\documentstyle[twocolumn,aps,psfig]{revtex}
\def\Sv{\vec{S}}
\def\Sz{S^z}
\def\Jp{{J'}}
\def\Jab{{\cal J}}

\def\Gap{{\sf E}}
\def\beq{\begin{equation}}
\def\eeq{\end{equation}}
\def\bea{\begin{eqnarray}}
\def\eea{\end{eqnarray}}
\def\nn{\nonumber}
\def\ie{{\it i.e.}}
\def\plus{\hbox{+}}
\font\amsmath=msbm10

\def\Zed{\hbox{\amsmath Z}}

\begin{document}
\tolerance 50000
\preprint{
\begin{minipage}[t]{1.8in}
\rightline{SISSA 93/98/EP}
\rightline{cond-mat/9808312}
\rightline{}
\end{minipage}
}

\draft

\twocolumn[\hsize\textwidth\columnwidth\hsize\csname @twocolumnfalse\endcsname

\title{A Strong-Coupling Approach to the Magnetization Process of Polymerized
       Quantum Spin Chains}

\author{A.\ Honecker}
\address{International School for Advanced Studies,
      Via Beirut 2-4, 34014 Trieste, Italy.}

\date{August 28, 1998}
\maketitle
\begin{abstract}
\begin{center}
\parbox{14cm}{
Polymerized quantum spin chains (\ie\ spin chains with a periodic
modulation of the coupling constants) exhibit plateaux in their
magnetization curves when subjected to homogeneous external magnetic
fields. We argue that the strong-coupling limit yields a simple
but general explanation for the appearance of plateaux as well as of
the associated quantization condition on the magnetization. We then
proceed to explicitly compute series for the plateau boundaries of
trimerized and quadrumerized spin-1/2 chains. The picture is completed
by a discussion how the universality classes associated to the transitions
at the boundaries of magnetization plateaux arise in many cases
from a first order strong-coupling effective Hamiltonian.
}
\end{center}
\end{abstract}

\vskip1pc]

\noindent
Quantum spin systems at low (or zero) temperatures can exhibit plateaux
in their magnetization curves when subjected to strong external fields.
Such phenomena in quasi-one-dimensional systems have recently
been the subject of intense interest. In one dimension, there is an intriguing
interplay between theoretical progress on a systematic understanding
of the underlying mechanisms (see e.g.\ \cite{OYA}) and an increasing
number of experiments (see e.g.\ \cite{NHSKNT,STKTUOTMG}) on materials
which are believed to be predominantly one-dimensional.

Here we study polymerized spin-$S$ quantum spin
chains in a magnetic field. Their Hamiltonian is given by
\beq
H = \sum_{x} J_x \Sv_x \Sv_{x+1} - h \sum_{x} \Sz_x \, ,
\label{hOp}
\eeq
where we assume periodicity of the coupling constants
with period $p$, \ie\
\beq
J_x = J_{x+p} \, .
\label{Jper}
\eeq
We will mostly concentrate on spin $S=1/2$ and the
antiferromagnetic regime $J_x \ge 0$. The zero-temperature
magnetization process of the $S=1/2$ polymerized chains (\ref{hOp})
was studied in \cite{CaGy} using finite-size diagonalization
and a perturbative bosonization analysis around the case of equal coupling
constants $J_x = J$ (apart from this and the dimerized case,
only some trimerized \cite{Hida,Okamoto} and quadrumerized \cite{ChHi}
cases seem to have been studied in the literature). Here we wish
to complete the picture by discussing the `strong-coupling' limit
where at least one coupling constant is small with respect to
the others, \ie\ $J_{x_0} \to 0$.

As is known e.g.\ from studies of spin-ladders \cite{CHPa,CHPb},
the magnetization process is easy to understand if some
$J_{x_0} = 0$. In this limit, the chain (\ref{hOp}) decouples into
clusters of $p$ spins. These `strongly coupled' clusters magnetize
independently such that at zero temperature the magnetization
$\langle M \rangle$ can only take finitely many values. For
spin $S$ they are subject to the quantization condition
\beq
p S \left(1 - \langle M \rangle \right) \in \Zed \, ,
\label{qCond}
\eeq
with a normalization such that the magnetization has
saturation values $\langle M \rangle = \pm 1$. This quantization
condition was obtained (for $S=1/2$) in \cite{CaGy} and is
a special case of a more general condition written down
in \cite{CHPb}. In particular the latter was also motivated
by considering a limit in which the system decouples into
clusters of finitely many spins. In fact, this counting
argument is completely independent of the internal coupling
inside the cluster of the $p$ spins. The quantization condition
(\ref{qCond}) is therefore insensitive to details of the
model. However, not only the transition values of
the magnetic field but also the question if a possible plateau
is realized even in this limit depends on the precise coupling
inside the cluster. For the linear arrangement (\ref{hOp})
and antiferromagnetic $J_x > 0$ ($x \ne x_0$), all values
of $\langle M \rangle$ permitted by (\ref{qCond}) are indeed
realized at $J_{x_0} = 0$.

Clearly, it remains also to
be shown that the quantization condition (\ref{qCond}) is
indeed valid at generic points in the parameter space,
not only for the special points where some $J_{x_0} = 0$.
This can be supported by series expansions
around the decoupling point, an issue to which we shall
return below.

A first property which one can derive for the full interacting
spin-$1/2$ system is the upper critical field $h_{\rm uc}$ at which
the transition to a fully polarized ferromagnetic state takes place.
For antiferromagnetic $J_x \ge 0$ it is simply
given by the vanishing of the gap for the one-spinwave dispersion
above the ferromagnetic background. The value of $h_{\rm uc}$ is
therefore given by the maximal eigenvalue of the following
$p \times p$ matrix
\beq
{1 \over 2}
\pmatrix{
J_p \plus J_1 & -J_1 & 0 & \cdots & 0 & -{\rm e}^{i k} J_p \cr
-J_1 & J_1 \plus J_2 & -J_2 & 0 & \cdots & 0 \cr
0 & -J_2 & \ddots & \ddots & \ddots & \vdots \cr
\vdots & \ddots & \ddots & \ddots & \ddots & 0 \cr
0 & \cdots & 0 & \ddots & \ddots & -J_{p-1} \cr
-{\rm e}^{-i k} J_p & 0 & \cdots & 0 & -J_{p-1} & J_{p-1} \plus J_p \cr
} \, ,
\label{dHuc}
\eeq
where $k$ arises from a Fourier transform using the periodicity
(\ref{Jper}) (see also \cite{Hida} for a detailed analysis of
a related special case). For antiferromagnetic $\Jp > 0$,
the lowest energy excitations
occur at $k=0$ for $p$ even and at $k=\pi$ for $p$ odd
if we introduce the momentum by a translation of $p$ sites
(as in (\ref{dHuc})).

In order not to get lost in too many parameters, we restrict
to the same subspace that was also considered in \cite{CaGy} before
we proceed further. We will now concentrate on the following
periodic arrangement of coupling constants
\beq
J_x = \cases{\Jp & for $x \in p\Zed$, \cr
              J   & otherwise. \cr}
\label{specCoupl}
\eeq
Now we return to the computation of the largest eigenvalue of
(\ref{dHuc}). The case $p=2$ is a bit special; the
correct specialization of (\ref{dHuc}) to $p=2$ reads (with
the notation (\ref{specCoupl})):
\beq
{1 \over 2}
\pmatrix{
\Jp + J & - J - {\rm e}^{i k} \Jp \cr
- J - {\rm e}^{-i k} \Jp & J + \Jp \cr
} \, .
\label{dHucP2}
\eeq
Using (\ref{dHucP2}) for $p=2$ and (\ref{dHuc}) for $p=3$ and $4$
at $k = p \pi$ (modulo $2 \pi$) we find
\bea
h_{\rm uc}^{(p=2)} &=& J + \Jp \, ,
\label{hUcP2} \\
h_{\rm uc}^{(p=3)} &=& {3 \over 4} J + {1 \over 2} \Jp 
        + {1 \over 4} \sqrt{9 J^2 - 4 J \Jp + 4 \Jp^2} \, ,
\label{hUcP3} \\
h_{\rm uc}^{(p=4)} &=& J + {1 \over 2} \left(\Jp 
             +\sqrt{2 J^2 - 2 J \Jp + \Jp^2} \right) \, ,
\label{hUcP4}
\eea
respectively.

Next we turn to series expansions of the plateau boundaries
for $p \le 4$. For the present systems, we expect that the sharp
steps between the magnetization plateaux which are present for
$\Jp = 0$ or $J=0$ soften as soon as one turns on $J, \Jp >0$,
but that nothing further happens. This scenario was in fact confirmed
by the numerical and perturbative analysis around $J = \Jp$ of
\cite{CaGy}.

For $p=2$ the only non-trivial plateau is located at
$\langle M \rangle = 0$. Its boundary is given by the $k=0$ spin
gap $\Gap^{(p=2)}$. Series expansions in $\Jp/J$ for this gap
have already been carried out some time ago in \cite{Harris}
up to third order and have recently been extended to nineth order
in \cite{BRT}. Adding a further order to eq.\ (29) of
\cite{BRT} (in passing we have also checked eqs.\ (28) and (30)
loc.cit.) one arrives at:
\bea
{\Gap^{(p=2)} \over J} &=&
1 - {1 \over 2} \Jab - {3 \over 8} \Jab^2
  + {1 \over 32} \Jab^3 - {5 \over 384} \Jab^4
  - {761 \over 12288} \Jab^5 \nn \\
&&+ {18997 \over 1769472} \Jab^6
  + {21739 \over 7077888} \Jab^7
  - {214359199 \over 6794772480} \Jab^8 \nn \\
&&+ {11960596181 \over 4892236185600} \Jab^9
  + {833277779047 \over 117413668454400} \Jab^{10} \nn \\
&&+ {\cal O}\left(\Jab^{11}\right) \, .
\label{hcP2}
\eea
Here we have used the abbreviation
\beq
\Jab = {\Jp \over J}
\label{Jab}
\eeq
in order to make the presentation more compact.

A few remarks may be in place regarding the method used here
which is summarized e.g.\ in section~3 of \cite{myself}.
Like the method of \cite{BRT} it exploits the fact that the
leading coefficients of the series can be obtained on
a finite lattice. However, we use recurrence relations
for the coefficients and an exact symbolic representation
throughout the computation while in \cite{BRT} a symbolic
result was reconstructed from a high-precision numerical
computation. Presumably, cluster expansion algorithms (see e.g.\
\cite{WKO}) would be more efficient than the two aforementioned
methods, but we prefer a simple-minded approach because of the
ease with which it can be applied to $p > 2$ as well. 

For $p=3$, there is a plateau at $\langle M \rangle = 1/3$, as one
infers from the above inspection of the case $\Jp=0$. Its lower and upper
boundaries ($h_{c_1}^{(p=3)}$ and $h_{c_2}^{(p=3)}$, respectively)
are determined by the $k=\pi$ gap of the single-spin
excitations. Up to fifth order in $\Jp$, one finds the following
series:
\bea
{h_{c_1}^{(p=3)} \over J} &=&
{8 \over 9} \Jab
 + {211 \over 810} \Jab^{2}
 - {77437 \over 1312200} \Jab^{3}
 + {7606883 \over 188956800} \Jab^{4} \nn \\
&& + {7188324510751 \over 269989034112000} \Jab^4
 + {\cal O}\left(\Jab^6\right) \, , \nn \\
{h_{c_2}^{(p=3)} \over J} &=&
{3 \over 2} - {1 \over 18} \Jab
 - {521 \over 6480} \Jab^{2}
 - {394169 \over 6998400} \Jab^{3} \nn \\
&& - {2260895171 \over 79361856000} \Jab^{4}
 - {535736196039221 \over 43198245457920000} \Jab^{5} \nn \\
&& + {\cal O}\left(\Jab^6\right) \, .
\label{hcP3}
\eea

Lastly, for $p=4$ all relevant excitations have $k=0$ in the
antiferromagnetic regime $J, \Jp > 0$. For reasons that should
be obvious from the results we content ourselves with second
order series for the gap $\Gap^{(p=4)}$ and the lower
and upper boundaries of the $\langle M \rangle = 1/2$
plateau ($h_{c_1}^{(p=4)}$ and $h_{c_2}^{(p=4)}$, respectively):
\bea
{\Gap^{(p=4)} \over J} &=& 
{1\over 2} \left(1+\sqrt{3}-\sqrt{2} \right)
- {1 \over 24} \left(4+\sqrt{6}+\sqrt{2}\right) \Jab \nn\\
&&- {1 \over 132480}
\bigl( 3079 \sqrt{6} - 163960 \sqrt{3} + 28775 \sqrt{2} \nn \\
&& \qquad + 276026 \bigr) \Jab^2 \nn
+ {\cal O}\left(\Jab^3\right) \, ,  \nn \\
{h_{c_1}^{(p=4)} \over J} &=&
{1\over 2} \left(1+\sqrt{3}-\sqrt{2} \right)
+ {2 \sqrt{6} + 8 \sqrt{2} + 17 \over 48} \Jab \nn \\
&&+ {1 \over 14837760}
\bigl( - 2385712 \sqrt{6} + 4730320 \sqrt{3} \nn \\
&& \qquad - 4947835 \sqrt{2} + 7747472
  \bigr) \Jab^2 + {\cal O}\left(\Jab^3\right) \, , \nn \\  
{h_{c_2}^{(p=4)} \over J} &=&
1 + {1 \over \sqrt{2}}
 + {2 \sqrt{2} - 3 \over 16} \Jab
 + {21691 \sqrt{2} - 31648 \over 43008} \Jab^2 \nn \\
&& + {\cal O}\left(\Jab^3\right) \, .
\label{hcP4} \\
\nn
\eea

With the choice of coupling constants (\ref{specCoupl})
there is a second decoupling limit, namely $J \to 0$, which
for $p \ge 3$ is not equivalent to the case discussed before.
This limit is special in that several of the coupling constants
(\ref{Jper}) vanish at the same time. This leads to $p-2$ free
spins in zeroth order in $J$. These free spins are immediately
polarized once a magnetic field is applied. Only the two
spins coupled by $\Jp$ require a finite magnetic field to
polarize. This leads to an $\langle M \rangle = 1 - 2/p$
plateau whose upper boundary is given by $h_{c_2}^{(p)}
= \Jp - {\cal O}(J)$.

At first order in $J$, one now has to perform degenerate
perturbation theory for the free spins. It turns out
that at this order they behave as isolated clusters of
$p-2$ spins.
The corresponding transition fields have been tabulated in
\cite{CHPb} and are indeed a reasonable first approximation
to the plateau boundaries for $3 \le p \le 6$
of \cite{CaGy} at large $\Jp$.

It is actually not difficult to obtain expansions in $J$ for
some plateau boundaries. Poor convergence is, however,
to be anticipated.
In the present case, internal properties of the decoupled clusters
are computed perturbatively (which were already taken care of
exactly at zeroth order in the expansions around $\Jp = 0$).
This is reflected e.g.\ in the fact that the fundamental excitations
start to disperse (\ie\ depend on $k$) only in the second
order in $J$.

At $p=3$ we find the following eleventh order series for the
boundaries of the $\langle M \rangle = 1/3$ plateau:
\bea
{h_{c_1}^{(p=3)} \over J} &=&
\Jab^{-1} + {3  \over 2} \Jab^{-2} - {107 \over 32} \Jab^{-4}
- {1185  \over 256} \Jab^{-5} + {845 \over 256} \Jab^{-6} \nn \\
&&+ {537329 \over 24576} \Jab^{-7}
+ {834121 \over 32768} \Jab^{-8}
- {310154551 \over 7077888} \Jab^{-9} \nn \\
&&- {15865989569 \over 84934656} \Jab^{-10}
+ {\cal O}\left(\Jab^{-11}\right) \, , \nn \\
{h_{c_2}^{(p=3)} \over J} &=&
\Jab + {1 \over 2} - {1 \over 4} \Jab^{-2}
-{5  \over 64} \Jab^{-3} + {19  \over 64} \Jab^{-4} \label{hcP3a}  \\
&& - {1317 \over 8192} \Jab^{-6} + {4199 \over 196608} \Jab^{-7}
+ {96157 \over 589824} \Jab^{-8} \nn \\
&& - {3539135 \over 28311552} \Jab^{-9}
- {133012373 \over 679477248} \Jab^{-10}
+ {\cal O}\left(\Jab^{-11}\right) \, .
\nn
\eea
This result is valid irrespective of the sign of $J$.
Indeed, we find agreement with the second-order result of
\cite{Hida} for ferromagnetic coupling $J < 0$.
For $J<0$ and $\Jp>0$, there is an experimental
realization of a trimerized system: 3CuCl$_2 \cdot$2dioxane.
However, since the coupling constants of this material
are roughly given by $J / \Jp \approx - 5$, the experimental
magnetization curve \cite{AAIAG} is far outside the range
of validity of our series (\ref{hcP3a}); actually no magnetization
plateau is observed experimentally.

For $p=4$ and antiferromagnetic $J, \Jp > 0$, one finds the following
counterpart of (\ref{hcP4}):
\bea
{\Gap^{(p=4)} \over J} &=&
1-{1 \over 4} \Jab^{-1}
+ {\cal O}\left(\Jab^{-2}\right) \, , \nn \\
{h_{c_1}^{(p=4)} \over J} &=&
1+{1 \over 2} \Jab^{-1}
+ {1 \over 4} \Jab^{-2}
+ {1 \over 4} \Jab^{-3}
- {5 \over 32} \Jab^{-4} \nn \\
&&- {41 \over 64} \Jab^{-5}
- {201 \over 256} \Jab^{-6}
- {497 \over 2048} \Jab^{-7}
+ {11887 \over 8192} \Jab^{-8} \nn \\
&& + {52929 \over 16384} \Jab^{-9}
+ {180845 \over 65536} \Jab^{-10}
+ {\cal O}\left(\Jab^{-11}\right) \, , \nn \\
{h_{c_2}^{(p=4)} \over J} &=&
\Jab + {1 \over 2}
+ {1 \over 4} \Jab^{-1}
- {1 \over 16} \Jab^{-3}
- {3 \over 16} \Jab^{-4} \nn \\
&& - {7 \over 512} \Jab^{-6}
+ {449 \over 4096} \Jab^{-7}
+ {715 \over 4096} \Jab^{-8} \nn \\
&&+ {6555 \over 65536} \Jab^{-9}
- {62051 \over 524288} \Jab^{-10}
 + {\cal O}\left(\Jab^{-11}\right) \, .
\label{hcP4a}
\eea
For the gap $\Gap^{(p=4)}$ we restrict to second order only, since
the high degeneracy at $\langle M \rangle = 0$ for $J = 0$ starts
to invalidate our approach at the third order.

\begin{figure}
\psfig{file=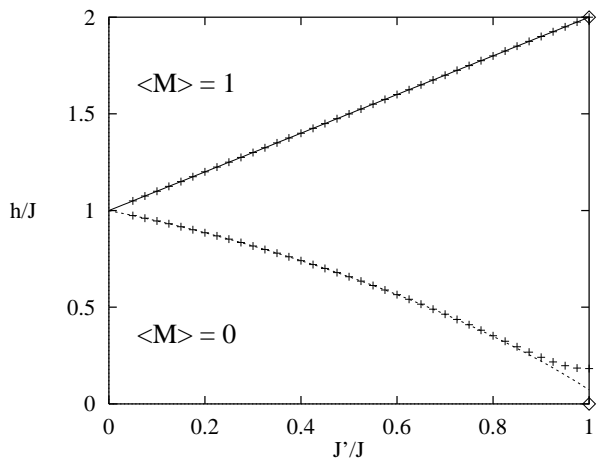,width=\columnwidth}
\smallskip
\caption{
Gap and transition fields for $p=2$.
The full line shows the upper critical field (\ref{hUcP2}).
The dotted line is our tenth order series expansion (\ref{hcP2})
for the spin gap. Crosses show the $L=24$
numerical data of Cabra and Grynberg and the diamonds denote
the magnetic fields $h$ at $\Jp = J$ associated to
$\langle M \rangle = 0$ and $\langle M \rangle = 1$
respectively.
}
\label{figP2}
\end{figure}

\begin{figure}
\psfig{file=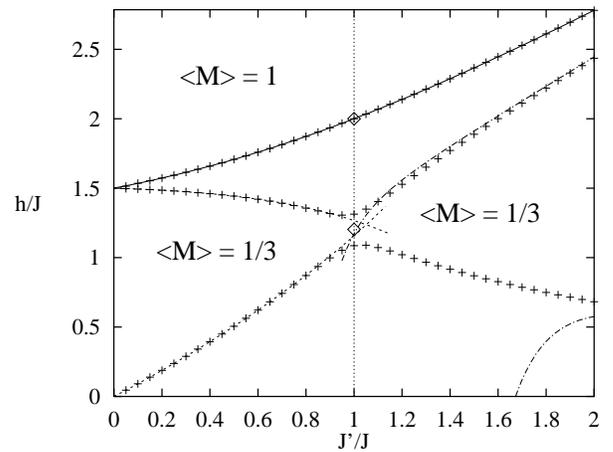,width=\columnwidth}
\smallskip
\caption{
Transition fields for $p=3$.
The full line shows the upper critical field (\ref{hUcP3}),
the dotted lines the series (\ref{hcP3}) for small $\Jp$
and the dashed-dotted lines the series (\ref{hcP3a})
for small $J$. Crosses show $L=24$
numerical data of Cabra and Grynberg and the diamonds denote
the magnetic fields $h$ at $\Jp = J$ associated to
$\langle M \rangle = 1/3$ and $\langle M \rangle = 1$
respectively.
}
\label{figP3}
\end{figure}

\begin{figure}
\psfig{file=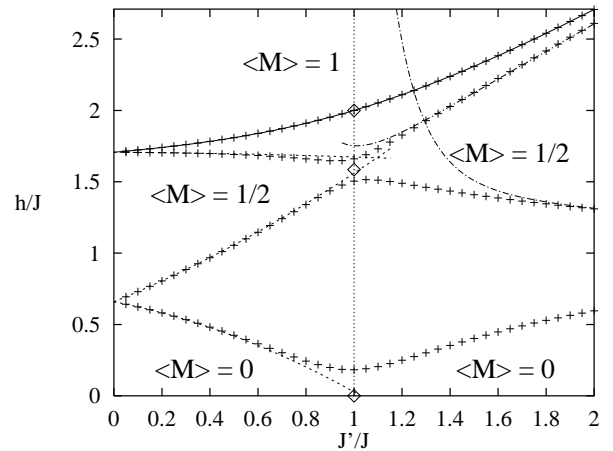,width=\columnwidth}
\smallskip
\caption{
Same as Fig.\ \ref{figP4}, but for $p=4$. The full line
shows (\ref{hUcP4}), the dotted lines (\ref{hcP4})
and the dashed-dotted lines (\ref{hcP4a}).
Diamonds denote magnetic fields associated to
$\langle M \rangle = 0$, $1/2$ and $1$, respectively.
}
\label{figP4}
\end{figure}

We compare our perturbative results to the $L=24$ numerical data \cite{CaGy}
in Figs.\ \ref{figP2}-\ref{figP4}. The full lines show our results for the
upper critical fields $h_{\rm uc}$, the dotted lines denote
the series expansions around $\Jp = 0$ while the dashed-dotted lines
in Figs.\ \ref{figP3},\ref{figP4} show the expansions around
$J=0$. Crosses denote $L=24$ numerical data of \cite{CaGy}
and the diamonds show the magnetic fields associated to the
plateau-values of $\langle M \rangle$ at $\Jp = J$.
For the isotropic chain they are $h=0$ for $\langle M \rangle = 0$
and $h = 2 J$ for $\langle M \rangle = 1$. The fields associated
to $\langle M \rangle = 1/3$ and $\langle M \rangle = 1/2$ are
computed from the Bethe-ansatz solution of the Heisenberg chain
(see e.g.\ \cite{CHPb}). Since Abelian bosonization predicts
the plateaux to open for $\Jp \ne J$ \cite{CaGy}, the diamonds
denote the expected ending points of the magnetization plateaux.

The case $p=2$ is shown for completeness in Fig.\ \ref{figP2}.
Since here the regimes $\Jp \le J$ and $\Jp \ge J$ are equivalent, we
display only the former. Here the dotted line shows our tenth
order series expansion (\ref{hcP2}) for the excitation energy at $k=0$.
The overall agreement is excellent, as has
already been observed for the gap in \cite{BRT}.

Fig.\ \ref{figP3} shows the next case, $p=3$. The series (\ref{hcP3})
are in excellent agreement with the $L=24$ numerical data for
$\Jp < J$. Even the ending point of the $\langle M \rangle = 1/3$
plateau is reproduced quite well. For $\Jp > J$, the upper boundary
of the plateaux is also reproduced reasonably by (\ref{hcP3a}),
while the agreement for the lower boundary is poor despite
the length of the series. This is not entirely surprising as
we have remarked above. In fact, inspection of the expression for
$h_{c_1}^{(p=3)}$ in (\ref{hcP3a}) shows that the coefficients
get larger with increasing order such that this series
might actually not converge in the region shown in Fig.\ \ref{figP3}.

This comparison of perturbation theory and finite-size
diagonalization is completed with the case $p=4$ in Fig.\ \ref{figP4}.
The expansions (\ref{hcP4}) around $\Jp = 0$ compare again
favourably with the numerical data of \cite{CaGy} although the
series are only of second order. Also the series for $h_{c_2}^{(p=4)}$
in (\ref{hcP4a}) agrees quite well with the numerical data for
$\Jp > J$, while that for $h_{c_1}^{(p=4)}$ yields good agreement
at least at the right boundary of Fig.\ \ref{figP4}. The small-$J$
series for $\Gap^{(p=4)}$ is not even shown, since due to its low order it
cannot be expected to give sensible results in the region of
interest. As in the case $p=3$, the limited quality of the series
for $\Jp > J$ can be expected on general grounds and is also
indicated by inspection of the value of the coefficients in
(\ref{hcP4a}). 

The comparison of the series (\ref{hcP3a}) and (\ref{hcP4a})
with the numerical data of \cite{CaGy} is complicated by the fact
that the latter does not extend into the region of small $J$ --
only the region $\Jp \ll J$ is covered well. We have therefore
performed some further numerical computations for $p=3$ and
$p=4$. At sufficiently small $J$ one can then nicely verify the
series order by order -- much in the spirit of \cite{BRT}. In
this manner we have verified the lowest five to
six orders of all series in (\ref{hcP3a}) and (\ref{hcP4a})
(a standard numerical accuracy is not sensitive to the highest
orders).

It has been pointed out recently by several authors (see e.g.\
\cite{Totsuka,Mila,TLPRS,Tot,CJYFHBLHP,FuZh}) that the strong-coupling
approach can be extended to describe the transitions between plateaux by an
effective Hamiltonian. For the case discussed here, one will
in general have to retain two states per site in first order
in $\Jp$. These two states correspond to the two plateau groundstates
at $\Jp=0$ between which we wish to describe the transition.
If the coupling constants are chosen to preserve parity
(as is the case e.g.\ for (\ref{specCoupl})), symmetry
arguments imply that the effective Hamiltonian is an XXZ-chain.
Hence one can immediately carry over some well-known universal
properties of the XXZ-chain (see e.g.\ section II of \cite{CHPb}
for a review) to polymerized spin chains. Firstly, the mapping
to the XXZ-chain in the strong-coupling limits implies that
the exponents of the correlation functions at the plateau boundaries
are given by
\beq
\eta_z = 2 \, , \qquad \eta_{xy} = {1 \over 2} \, .
\label{valExpo}
\eeq
Secondly, the transitions at the plateau boundaries are predicted
to be of the DN-PT type \cite{DzNe,PoTa}, \ie\ the magnetization
as a function of applied field $h$ has a square-root behaviour
close to the plateau boundaries. The same conclusions are obtained by the
Abelian bosonization analysis of the limit $\Jp \to J$ \cite{CaGy}.
This follows from results in the theory of commensurate-incommensurate
transitions \cite{SchulzCI} which in addition imply that the exponents
(\ref{valExpo}) as well as the DN-PT square-root behaviour should
be universal. The fact that identical conclusions are reached by
considering two different limiting cases are in agreement with such
a universal scenario.

Before concluding, it should be mentioned that the quantization
condition (\ref{qCond}) may have to be relaxed in certain
cases. For example, it has been shown (see e.g.\
\cite{Totsuka,TNK}) that an $\langle M \rangle = 1/2$ plateau
can appear if a next-nearest neighbour interaction is added to
the dimerized chain, (\ref{hOp}) with $p=2$ (for generalizations
of this situation see \cite{FGKMW}). This phenomenon can be also
understood within the strong-coupling analysis \cite{Totsuka,Mila,TLPRS,Tot}
if one goes to first order, \ie\ beyond the na\"{\i}ve decoupling limit
$\Jp = 0$. The crucial r\^ole is played by the XXZ anisotropy
appearing in the first-order effective XXZ chain.
If this effective XXZ anisotropy turns out to be sufficiently
large ($\Delta > 1$), a gap opens and translational symmetry is
spontaneously broken. In this manner one finds a further
plateau precisely in the middle between the two values of
$\langle M \rangle$ predicted by considering just the limit $\Jp = 0$. 
This illustrates that
$p$ in (\ref{qCond}) should be taken as the period of the
groundstate which in general can be different (\ie\ an integer
multiple) of the period of the Hamiltonian.

To summarize, we have shown that the study of the strong-coupling
limit not only provides a simple way to understand the
magnetization process of polymerized spin chains qualitatively,
but that also quantitatively competitive results can be obtained with
moderate effort. In some respects, the situation is even nicer
than for spin ladders \cite{CHPa,CHPb}: The bare series in $\Jp$
yield good results in the entire region $\Jp < J$, including
the ending-points of the plateaux. Such favourable conditions
are probably a special feature of polymerized spin-1/2 chains, as
is the fact \cite{CaGy} that here the condition (\ref{qCond})
is necessary {\it and} sufficient for the appearance of a plateau
at $\Jp \ne J$.

It is straightforward to extend the approach of this paper
to more general interactions or to the computation of other quantities.
We are confident that further explicit strong-coupling computations
will provide a useful tool e.g.\ if new experimental data
is to be explained in terms of model Hamiltonians.

\medskip

I would like to thank D.C.\ Cabra and M.D.\ Grynberg for providing me with
their numerical data. I am grateful to them and P.\ Pujol for useful
discussions and comments.
This work was carried under under financial support with the
TMR grant FMRX-CT96-0012.

\end{document}